# Towards Data-Driven Precision Agriculture using Open Data and Open Source Software

**Jacob Høxbroe Jeppesen [a,\*], Rune Hylsberg Jacobsen [a],**
**Rasmus Nyholm Jørgensen [a], Thomas Skjødeberg Toftegaard [a]**

[a] Department of Engineering, Aarhus University, Aarhus, Denmark
* Corresponding author. Email: jhj@eng.au.dk

### Abstract

Information and communications technology (ICT) within the agricultural sector is characterized by a widespread use of proprietary data formats, a strong lack of interoperability standards, and a tight connection to specific hardware implementations resulting from vendor lock-in. This partly explains why ICT has not yet had its full impact within the domain. By utilizing the vast amount of publicly available open data, ranging from topographic maps to multispectral satellite images, the economically and environmentally optimal farming practices can be advanced beyond state of the art. This paper addresses the potential of applying publicly available information sources to improve crop production, with emphasis on yield optimization. This potential is evaluated based on free public data for the growth season 2016 by examining winter wheat production for a selected region in Denmark. Data aggregation is performed by promoting open-source software tools as a foundation for decision support. That allows the farmer, or another domain expert, to query a certain crop type, merge this information with other data sets, and perform analysis on data ranging from sub-field analysis to statistics on national/regional scale. The registration of field polygons and sowed crop types for fields in Denmark, alongside with detailed geographic data and free satellite images, enable us to exploit publicly available data of high quality, which can be applied to perform further analysis.

This work presents results from the Future Cropping partnership, aiming to improve crop production in Denmark. The direct benefits of employing a data infrastructure exhibiting a high degree of flexibility are addressed, alongside with a discussion on the importance.

## 1. Introduction

The agricultural sector occupies roughly 38% of the surface area of the Earth, with large parts of the rest being unsuitable for farming, such as deserts, mountains, and cities (Foley et al., 2011). From 1985 to 2005, this area has only increased by 2.4%, and the extra area was mostly gained through deforestation in tropical parts of the world, which is a non-viable solution (Foley et al., 2011). The food demand is expected to double from 2005 to 2050 (Tilman et al., 2011), hence the yield of the current fields must be increased. By bringing the yield within 95% of the potential of the field for 16 types of crops, it is estimated that an increase of 58% of food production can be obtained (Foley et al., 2011). One practice to facilitate this is the use of data-driven precision agriculture to take intra-field variations into consideration.

A requirement for precision agriculture is access to data, and in recent years, open data have continued to become more easily available. In particular, organized governmental data of high potential value is accessible from several countries (Harris & Baumann, 2014). The data themselves and the accessibility to them does not automatically create value however, they must be processed to obtain information and knowledge. To facilitate this, open source software packages are progressing tremendously, and have recently become sophisticated enough to match their proprietary counterparts (Steiniger & Bocher, 2009).

Information and communication technology (ICT) within the agricultural sector for obtaining site-specific farming has had a slow adoption rate, partly due to the lack of standardization and partly because it required a shift from field-level data to a Geographic Information System (GIS) infrastructure (Nikkilä et al., 2010). Open source software solutions have progressed tremendously in recent years and have been complemented by open data of high quality, such as the Sentinel 2 satellite images, providing 12 spectral bands in resolutions down to 10 m × 10 m. Furthermore, the agricultural domain is focusing on standardizing the data, making it feasible to, e.g., collect data from machinery and merge it with satellite data and topographic maps for decision support (Nash et al., 2009).

The Future Cropping partnership is a Danish research project with the goal of advancing the agricultural sector beyond current state of the art. As a collaboration between two universities and eleven companies, it aims to close the gap between academia and industry, with special emphasis on the opportunities in data-driven agriculture and precision farming. A clear goal is to ease the process for the farmer to apply site-specific farming, by investigating methods to improve data communication between involved instances, ranging from consultants and farmers to big data algorithms and machinery.

Furthermore, data mining techniques will be used to extract valuable information from the large amounts of data within the domain. This enables the agricultural sector to improve yield while minimizing the environmental impact.

This paper investigates the use of open data for precision agriculture, and demonstrates a proof of concept implementation of a GIS toolkit providing a web interface and analysis capabilities, entirely based on open source software. The toolkit facilitates the use of open data sources, making it possible to perform decision support to complement existing farm management information systems. Two use cases are presented to evaluate the potential of the toolkit: one for sub-field analysis and one for statistics on a regional scale.

The rest of this paper is organized as follows: First, we describe the methods including the open data and open source software in Section II. In Section III, the results will be presented and discussed through two use cases, before Section IV concludes the paper and provides comments on future work.

## 2. Materials and Methods

Our work applies GeoNode for data management and QGIS for data analysis. The open data were comprised of field polygons, topographic maps, satellite images, and orthophotos.

GeoNode is an open geospatial content management system based on mature open source software packages, such as GeoExplorer to provide GIS data in a web application, GeoServer as a data manager using GeoNetwork for catalogization, and PostgreSQL with a PostGIS extension as database. The setup can be used for uploading/downloading data, browsing, sharing, and managing GIS data. It employs a user-based structure, with the capability of authentication requirements for specific data. In Figure 1, an overview of the included software packages is presented.

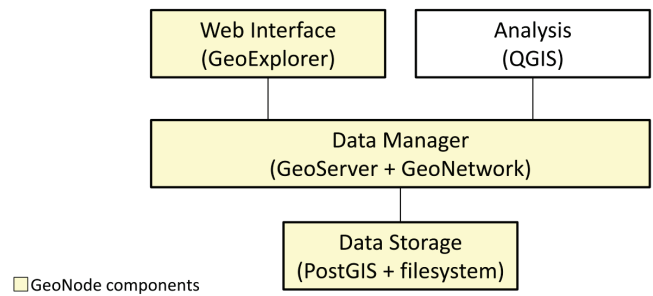

Figure 1. Overview of the open source software used in this paper.

The setup in Figure 1 is highly flexible, and can be modified according to user needs. GeoNode is deployed on a virtualized server infrastructure that can be scaled horizontally as the demand for processing power and memory increases. Users can upload and share measurement or machine data in Shape or GeoTIFF formats, for central storage and for sharing it with, e.g., consultants/agronomists. It is based on standardized formats and interfaces, and the data manager, most importantly GeoServer, supports standard geospatial protocols, such as the Web Map Service (WMS) and the Web Feature Service (WFS). This allows for easy querying of data by 3$^{rd}$ party products.

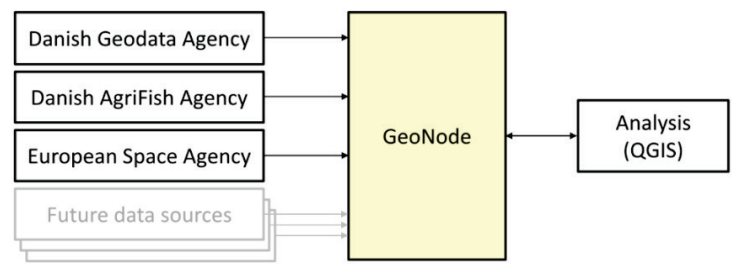

Figure 2. Overview of the data flow.

Figure 2 shows the data flow in the setup. The topographic maps and the orthophotos were provided by The Danish Geodata Agency, and the field polygons were provided by The Danish AgriFish Agency. The satellite images were provided by The European Space Agency (ESA), and comprises of multispectral images from the Sentinel 2 satellite.

The Normalized Difference Vegetation Index (NDVI) was calculated from the following expression provided by ESA (2016):

$$NDVI = \frac{NIR - VIR}{NIR + VIR} = \frac{Band\ 8 - Band\ 4}{Band\ 8 + Band\ 4}, \qquad (1)$$

where *NIR* and *VIR* are the near-infrared regions and the visible red regions respectively. The resolution is 10 m × 10 m.

QGIS was used to analyze and process the data with its multitude of plugins, alongside with its seamless integration with the programming language R for statistics purposes. QGIS runs on a regular workstation and connects to the data via the GeoServer. The processed data were then uploaded to GeoNode, such that all raw and processed data were stored centrally, with easy access to create maps and investigate results through the web interface, or to download the desired data. It should be noted that the entire system can also be hosted on a local server to ensure data privacy.

The open source toolkit presented in this paper follows a simple installation procedure, and provides possibilities of cloud computing for storage, sharing and visualization. GeoNode was customized to form the *"Future Cropping Data Platform"*, as seen in Figure 3, and was setup on a local server with public access.

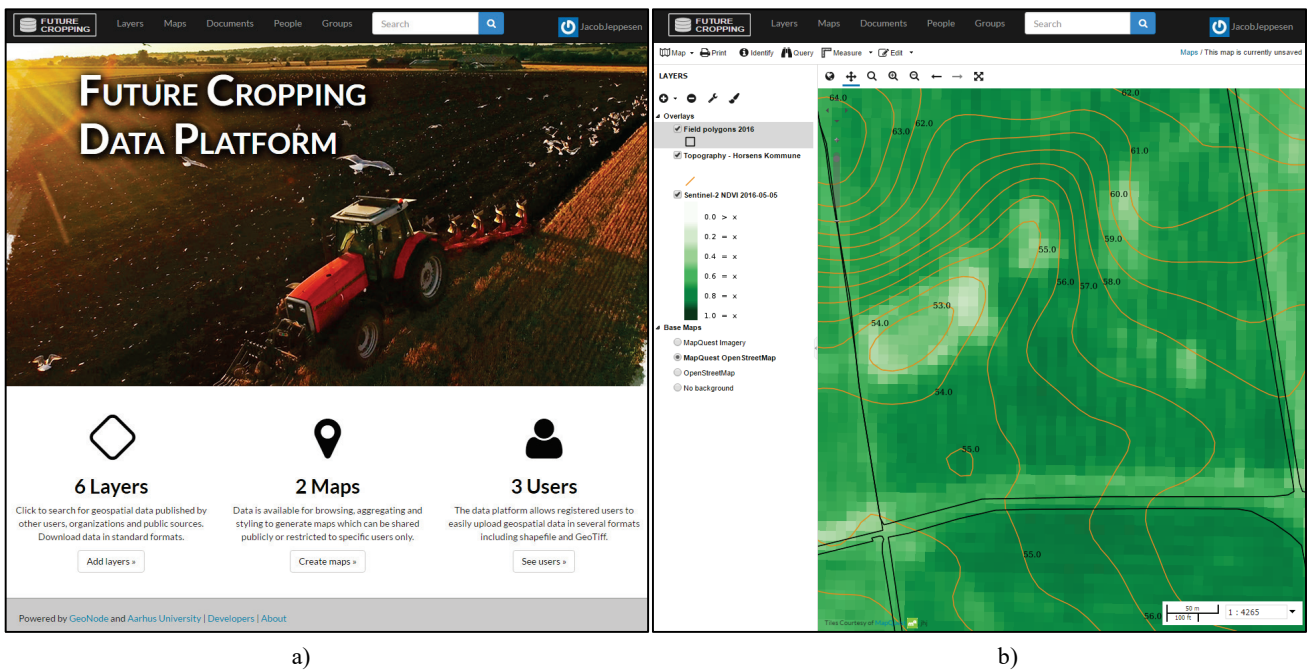

a)                                                                 b)

Figure 3. The web-based data platform – a) shows the front page and b) shows an online visualization of a map constructed from an NDVI map, a topographic map, and field polygons.

## 3. Results and Discussion

To evaluate the potential of the open source toolkit and data, we focused on two use cases. Case 1 was for the farmer and agronomist to easily access the farmer's data, and to be able to investigate his crops from satellite images and topographic maps. Case 2 was for researchers to perform statistics on a regional or national/regional scale, and demonstrate how the data can be applied for further analysis.

### 3.1. Use case 1: Sub-field scale analysis

The ability of assessing the status of crops remotely and compare it to previous years, results in several possibilities for the farmers and consultants/agronomists. By investigating NDVI maps and topographic maps, explanations for crops performing poorly can be investigated. In Figure 4, we investigate the potential of using the NDVI map from the Sentinel 2 satellite to assess the status of a field.

By overlaying the topographic map with the NDVI map, it is possible to find explanations for poor-performing patches. More importantly, it provides information of the spatial extent of the poor-performing patches, leading to a better overview of the potential economic impact. This aids in the decision of re-establishment of crops. Furthermore, the NDVI map can be used to optimize fertilization, and to calculate a site-specific fertilization map, which can be uploaded to the machinery.

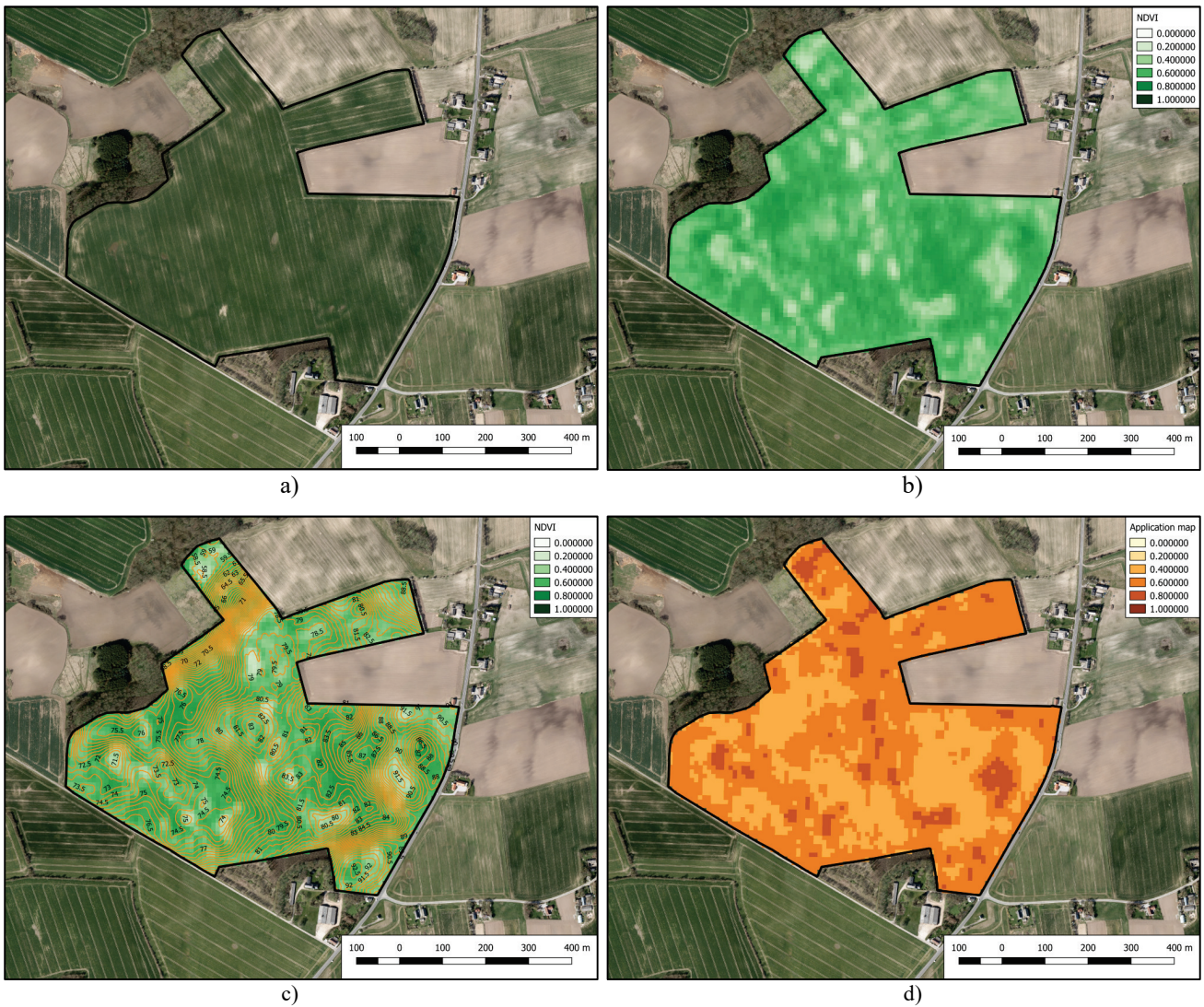

Figure 4. The analysis of a winter wheat field using Satellite imagery from 5 May 2016 - a) shows the orthophoto (from 2015) and field polygon, b) shows NDVI map of the field from the Sentinel 2 satellite, c) shows the field with the NDVI map and a topographic map, and d) shows an example of an application map based on the NDVI map.

Further possibilities were investigated by Seelan et al. (2003), where multiple use cases were tested from 1998 to 2001, and the economic gain directly coupled to remote sensing was investigated. One use case in the paper is the variable-rate nitrogen application for a 30 ha sugar beet field. This use case showed a total saving of US$7326 (US$244 per ha) and a 35% reduction of the amount of applied nitrogen. The variable-rate zones where determined from an NDVI map based on images from the IKONOS satellite. However, obstacles such as expensive image-processing and lack of standardization were met.

In Atzberger (2013), it is stated that access to data is too complicated in the agricultural sector, and that data should be easily visualized and downloaded. This obstacle can be circumvented by employing standards from the GIS community. Open source Spatial Data Infrastructures (SDIs), such as GeoNode, can provide a platform for data-sharing and visualization along with user authentication through a web interface. Furthermore, the support of standardized formats such as GeoTIFF for raster data and Shape files for vector data is an important factor to ensure interoperability with other software packages.

3.2. Use case 2: Regional scale analysis

In Figure 5, we queried winter wheat fields and merged satellite data on the polygons. The mean NDVI value for each field was calculated using the Zonal Statistics plugin in QGIS, and used to highlight fields performing poorly.

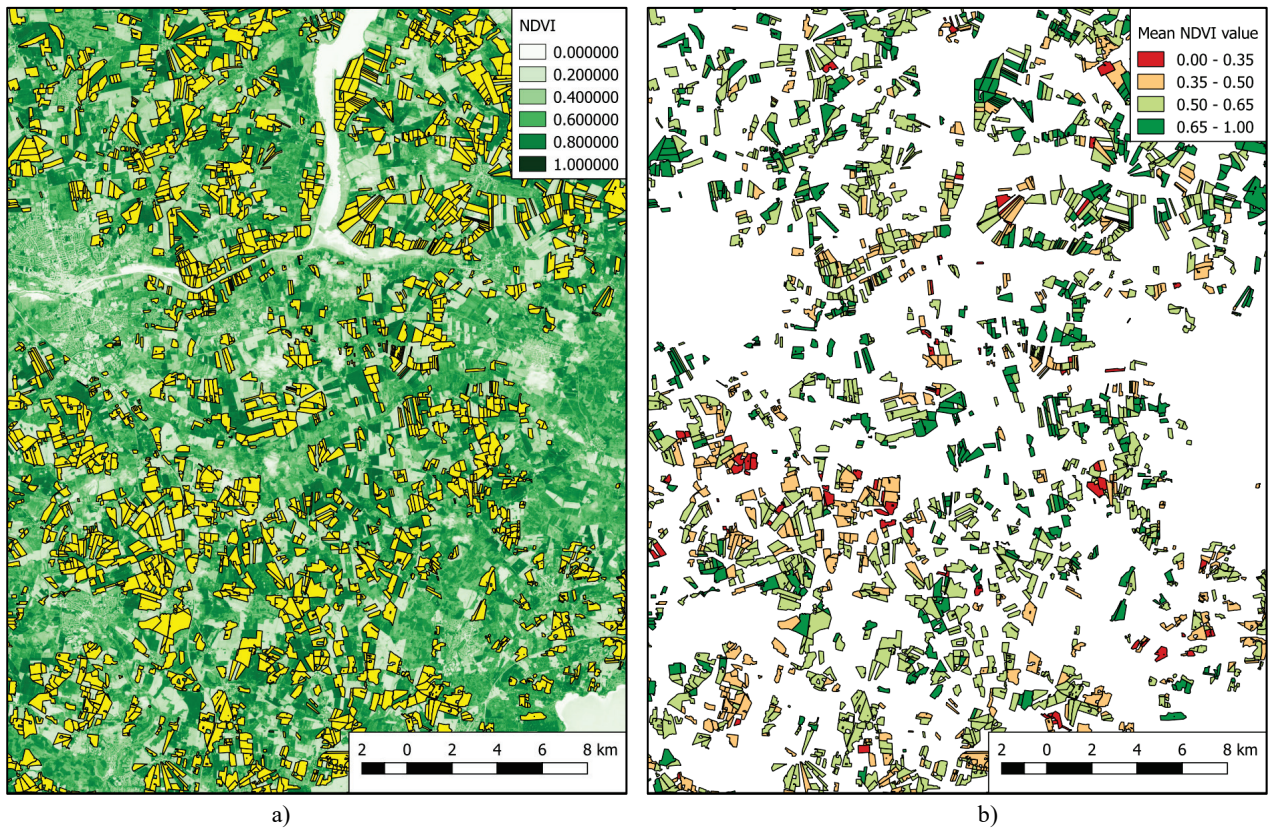

a)                                              b)

Figure 5. Regional scale analysis based on Sentinel 2 satellite imagery on 5 May 2016 – a) shows winter wheat fields (yellow) on top of an NDVI map and b) shows these fields categorized after their mean NDVI values.

By having the data in a standardized format, proper interoperability between software packages is ensured, and several analysis possibilities becomes easily available. The histograms in Figure 6 were made by running R scripts in QGIS. These show the possibilities of performing regional and even national scale analysis. They provide a tool to benchmark fields against each other, and grants a valuable overview of the status of the fields. Furthermore, the analysis was carried out on a regular workstation.

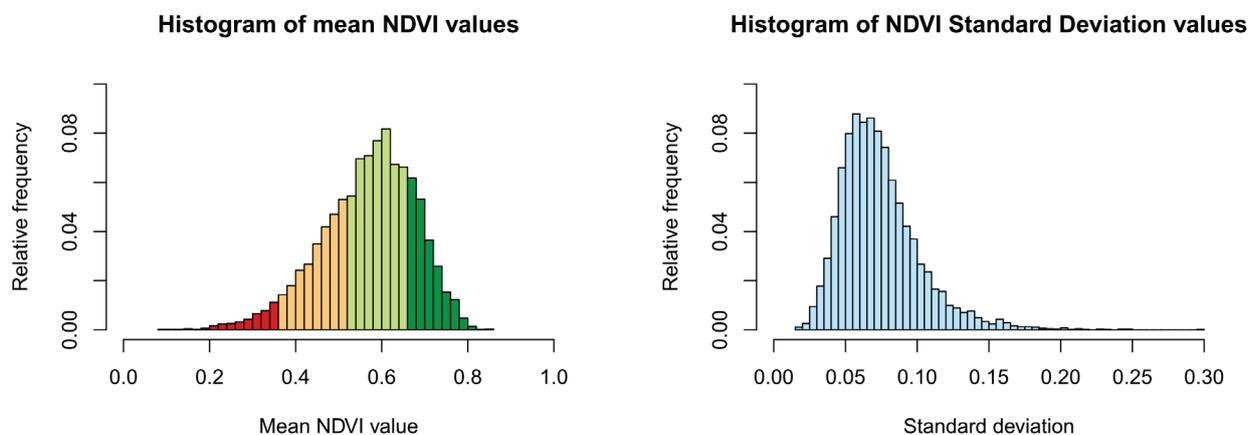

Figure 6. Histograms of the mean NDVI values and NDVI standard deviation values within fields for 6524 winter wheat fields on 5 May 2016 in Jutland, Denmark.

The value of performing these types of statistics is to draw conclusions on the general status of the fields, following, e.g., a harsh winter. This is of high relevance for the entire value chain, from the seed suppliers to determine whether a large quantity of crops will be re-established, to farmers determining the crop yield and banks to assess the economic status of the current growing season.

It should be noted that the NDVI maps and the statistics based on them in this paper do incorporate a certain degree of uncertainty. Issues such as atmospheric disturbances are not considered, and other vegetation indices might prove to be superior regarding the Sentinel 2 satellite (Frampton et al., 2013). The analytical strength of the presented setup, however, lies in its flexibility. It provides a way to quickly calculate the desired vegetation index, easily perform various calculations based on these, and to carry out a broad range of statistical analysis in a straightforward manner.

## 4. Conclusions

We have demonstrated a proof of concept toolkit, where a data platform was implemented by using the GeoNode framework and complementing this with QGIS as an analysis tool. The data platform grants the user access to a simple interface for data sharing based on a GIS web interface. It employs mature software packages and is designed to be easily modified and extended. Topographic maps, field polygons, and satellite images were collected in the data platform. These data can be basis for decision support on a sub-field scale, and can be supplemented by machine and sensor data from the farmer. Furthermore, analysis and statistics on a regional or even national scale can be performed using only open data and open source software.

Several directions for future work are feasible, ranging from data communication and ICT infrastructure to data analysis of open data, machine data, and remote sensing data. In particular, data analysis based on the ESA Copernicus program encompassing the Sentinel satellites, complemented by a standardized GIS platform to ensure a seamless data connection from satellite to machinery, seems very promising

### Acknowledgements

This work was funded by Innovation Fund Denmark as part of the Future Cropping project (J. nr. 5107-00002B).